\documentclass[12pt]{iopart}
\usepackage{graphicx}
\usepackage{todonotes}
\usepackage{hyperref}
\usepackage{epigraph}
\usepackage{comment}
\RequirePackage{xcolor}
\definecolor{green}{rgb}{0,.9,.15}
\definecolor{violet}{rgb}{.8,.12,.86}

\begin{document}
\raggedright
\title[Analytic Gravitational Waves from Binary Black Holes in Elliptic Orbits]{Eccentric Pairs: Analytic Gravitational Waves from Binary Black Holes in Elliptic Orbits}
\author{Dillon Buskirk \footnote{Present address: Department of Physics \& Astronomy, University of Kentucky, Lexington, KY 40506, USA} and Maria C. Babiuc Hamilton \footnote{Corresponding author: \href{mailto:babiuc@marshall.edu}{babiuc@marshall.edu}}}
\address{Department of Physics, Marshall University, Huntington, WV 25755, USA}
\vspace{10pt}
\begin{indented}
\item[]
\end{indented}
\epigraph{If you thought that science was certain - well, that is just an error on your part.}{Richard P. Feynman}“Not only is the Universe stranger than we think, it is stranger than we can think.”
\begin{abstract}
Gravitational waves (GW) from eccentric binaries have intricate signals encoding important features about the location, creation and evolution of the sources. Eccentricity shortens the merger time, making the emitted GW statistically predominant in the observed data once detectors will reach the required sensitivity.
We present a novel implementation of fully analytical GW templates from eccentric binary black hole (BBH) mergers within the \texttt{Wolfram Mathematica} software. We increase the accuracy by identifying and minimizing the possible source of errors. 
We start with an overview of the physics involved in eccentric mergers, then assemble the strain for the inspiral by employing up to six post-Newtonian (PN) corrections. 
We complete the eccentric inspiral with the quasi-circular Backwards one Body (BoB) merger model in frequency, amplitude and phase, then we build the hybrid GW strain for the whole evolution of the binary. For low eccentricity we reach coincidence in the overlap, with no ambiguity in the time interval, a remarkable improvement from the usual matching techniques. 
For high-eccentricity we compensate for the implicit quasi-circular assumption of the BOB approach, by introducing a small rescaling in amplitude.
Our streamlined implementation is relevant for the new field of GW astronomy and is straightforward to understand, use and extend, offering researchers in the field a valuable open resource tool. 
\end{abstract}
%
%
%
%
%

\section{Introduction}
\label{sec:intro}

In the 17th century, Johannes Kepler introduced elliptical orbits and his three famous laws of planetary motion. Since then, it is universally accepted that all binary systems move in elliptical orbits around the center of mass. 
This empirical observation was used by Newton to build the law of universal gravitation, unchallenged for three centuries, until Einstein introduced the theory of general relativity (GR) in 1915. 
This theory explained the observed slight precession in the elliptical orbit of Mercury, known as the relativistic perihelion advance. 
More important, it showed that a binary is bound to spiral inward and circularize by releasing orbital energy as gravitational waves (GW) until the inevitable collision occurs.
It may seem like a rare scenario that two compact objects lock each other into a collapsing orbit. However, stars are usually born in pair, and more than $70\%$ of their population is massive enough to end up either as a black hole or a neutron star \cite{arXiv:1807.11489}. 

By the time they reach Earth, GWs have extremely small amplitudes, with an expected strain on a given length of about $10^{-20}$. 
This makes their detection very challenging and possible only for very strong gravity, as in the case of the two merging black holes first detected by the advanced Laser Interferometer Gravitational-Wave Observatory (aLIGO) in September 2015 \cite{arXiv:1602.03837}.
This discovery created a new branch of science, the Gravitational-wave Astronomy \cite{arXiv:gr-qc/9911034}.
The data and software employed in those detections are freely available \cite{arXiv:1912.11716}, enabling peer scientists to test and replicate them \cite{arXiv:2010.07244}, an essential requirement for the upholding of high scientific standards.

All the detections reported until now come from circularized binaries, but this is not necessarily because they had zero eccentricity \cite{arXiv:1912.05464}. 
The emission of GWs decreases the orbital eccentricity \cite{Peters:1964}, thus very small eccentricity is expected close to the merger, when the signal enters the sensitivity band of the current detectors, especially for isolated compact binaries evolving for a long time.
However, in certain populations of merging binaries living in dense stellar environments, such as galactic cores and globular clusters \cite{arXiv:1705.05848,arXiv:1606.04889}, or in isolated triplets \cite{arXiv:1703.06614}, eccentricity is present even at small separations. 
Current ground-based detectors such as aLIGO and aVirgo, did not reach yet the sensitivity levels required for the detection of eccentricity.
They are only tuned to observe the brightest GW sources, in a limited frequency band, very close to collision \cite{arXiv:1810.03521}. 
In the near future though, as the technical difficulties are solved, detection of eccentricities will become possible \cite{arXiv:1312.2953}. 
Moreover, next generation of gravitational-wave observatories are being built and will come online within the next decade \cite{arXiv:2012.03608, arXiv:1912.02622, arXiv:1907.11305, arXiv:2001.09793, arXiv:1802.06977, arXiv:2008.10332}.
Those instruments will be able to detect GWs with up to 10 times increase in sensitivity, to discern signals at an earlier stage in the life of the binary, when orbital eccentricity is still large and to track their evolution for a long time. 
It is thus only a matter of time until eccentric GWs will be detected. 

GWs with eccentricity are much richer in structure and they reveal valuable information on how the binary came together, the medium in which the pair resides and its long-time evolution \cite{arXiv:1806.05350, arXiv:1705.10781, arXiv:1606.09558}.
It is known that eccentricity decreases the merger time, meaning that GWs with eccentricity encoded in the signal will become ubiquitous once detectors will have the demanded sensitivity.
But to decipher any such detection and to determine their proprieties, it is essential to have accurate templates that are valid for small to moderate eccentricities.

GW templates are generated mainly using two families of models currently implemented in the LIGO Scientific Collaboration Algorithm Library (LALSuite) [LAL], the Effective-one-Body approach (EOB) \cite{arXiv:9811091, arXiv:1611.03703, arXiv:1806.01772} and the phenomenological frequency domain model (IMRPhenom) with its flavors \cite{arXiv:1005.3306, arXiv:2004.08302, arXiv:2001.10914, arXiv:2001.11412}.
Those waveform models are following in essence the same two-steps procedure to construct complete templates for the signal, from the time it enters the detector's frequency band, through the peak of the amplitude (or {\em chirp}) emitted when the two black holes collide, until it gradually becomes smaller, and ending when the final black hole settles down.
First, the binary motion is split in three distinct regions: {\em inspiral}, {\em merger}, and {\em ringdown}, then different mathematical formalisms are applied to obtain the waveform for each region. 
Finally, a complete GW template is generated for the whole evolution of the binary, by matching those regions in time or in frequency.   

The inspiral - or the weak field region - is modeled using different versions of the post-Newtonian (PN) formalism, which obtains solutions to Einstein's equations of GR as higher-order expansions of Newton's law of gravity in terms of the small PN parameter $x_{PN}=v^2/c^2$, also called the  \emph{slowness}.
Here, $v$ is the orbital velocity and $c$ the speed of light (see \cite{arXiv:9806108, arXiv:0907.3596, arXiv:1102.5192, arXiv:1310.1528, arXiv:1902.09801} for details on this analytical technique).

In the second region - during the merger - the orbital speed of the binary gets closer to the speed of light, rendering  the PN approximation invalid.
Here the dynamics of the binary is modeled by numerically solving Einstein's equations of GR on supercomputers and extrapolating at larger distance to calculate the resulting waveform (see \cite{BaumgarteNR} for an excellent review).
Informed by numerical relativity (NR) \cite{arXiv:0610122, arXiv:0706.1305, arXiv:0710.0158, arXiv:1005.3306, arXiv:1502.04953, arXiv:1810.10585}, the analytical models are extended to envelop the nonlinear physics of the merger, by using either the empirical IMRPhenom approach \cite{arXiv:0909.2867, arXiv:1611.00332}, or the EOB reformulation the PN theory \cite{arXiv:0704.1964, arXiv:0706.3732, arXiv:1607.05661}.
In the third region, the new black hole formed after collision settles down by ringing out the extra energy in form of quasi-normal modes radiation. 
Here analytical models are valid again and the GWs are calculated with the black hole perturbation theory \cite{arXiv:9712057, arXiv:9909058,  arXiv:0306120, arXiv:0905.2975, arXiv:1406.5983}.

The state-of-art methods mentioned above are continuously improved, driven by the increasing demands of the GW detection. 
However, most of them are limited to quasi-circular orbits (for good reviews see \cite{arXiv:1408.5505, arXiv:2012.01350}).
The LALSuite data analysis libraries has only one routine for generating eccentric GW templates based on \cite{arXiv:1602.03081}, and current analytical models including eccentricity \cite{arXiv:1602.03081, arXiv:1709.02007,  arXiv:2002.03285} are under development to increase the accuracy and include all the complex physics of elliptic binaries \cite{arXiv:1901.07038, arXiv:1909.11011}. 

Motivated by the impending detection of GWs from eccentric binaries, we align our present work with the demand for eccentric waveforms and focus on accurately implementing fully analytical, easily reproducible GW templates from eccentric binary black hole (BBH) mergers. 
We construct a series of new, fully analytical notebooks within the \texttt{Wolfram Mathematica} software for calculating the complete GW emitted by compact binaries in eccentric orbits, by starting with their inspiral, following through the highly nonlinear regime of the coalescence and ending with the ringdown phase.
We describe in detail how the entire template is built, by bringing together valuable information from research papers that focus only on specific regions of the orbit and exhibiting a high level of complexity hidden in scientific jargon, which makes the problem hard to understand.  
We open this black box and bring its intricate content to an approachable level, in a clear way, keeping the balance with the scientific rigor. Our goal is for the reader to understand, follow and reproduce our results, even without formal training in GW research. 

Another novelty of our implementation resides in the construction of the hybrid GW strain for the whole evolution of the binary, matching the eccentric inspiral with the analytic quasi-circular Backward-one-Body (BoB) model \cite{arXiv:1810.00040} for the merger, by completing them in frequency, amplitude and phase. 
We find that for low eccentricity we can connect the two waveforms with no ambiguity in the time interval, result relevant for the production of accurate GW waveforms from eccentric binaries.
Our research is applicable to the field of GW astronomy and offers researchers in the field a new tool for generating eccentric GW, as well as a blueprint for understanding, reproducing and extending our results, in order to reveal the physics of this kind of sources, in preparation for their imminent detection.

We start our presentation with a review of the Keplerian problem for eccentric binaries and its extension to the post-Keplerian parametrization in \Sref{sec:weakfield}. We clarify the complex physical and mathematical concepts required to solve analytically the problem of motion when eccentricity is intertwined with strong gravity, a topic often either overlooked, or hidden in heavy technical language. Our choice of solving the post-Newtonian Kepler's equation (based on \cite{arXiv:1707.02088}) is less used and has the advantage that is purely analytical. We finish this section by outlining the steps followed to assemble the analytic gravitational waveform during the inspiral. 
We continue in \Sref{sec:strongfield} by exposing first the difficulty in analytically determining the location of the transition between the inspiral and merger, then we follow by summarizing the important formulas employed by the merger model to arrive at the GW strain.

Once the theoretical  framework is laid out clearly, it is time to detail our new implementation and to explain how we built the complete waveform in \Sref{sec:complete}. 
 We start by establishing the time bounds for the binary evolution, then we implement the inspiral waveform. 
Lastly, we generate the complete GW template for the whole evolution of the binary, by matching the GW obtained for the inspiral with the merger GW in amplitude, frequency and phase, at a fixed transition time. 
We employ first small eccentricity, then we push the boundaries by devising a hybrid waveform for large eccentricity. 
We conclude in \Sref{sec:conclude} with a summary of our findings, their implications for the field of gravitational wave astrophysics, and suggest extensions. 

We work in geometrized units with $G = c = 1$, thus expressing time, space and mass function the mass of the binary system at infinity $M = M(\texttt{kg})/M_\odot(\texttt{kg})$ where $M_\odot$ is the mass of the sun.  
This allows an easy rescaling of our results for a given total mass. 
The physical units are recovered by multiplying mass with $M_\odot (\texttt{kg})= 1.98892 \times 10^{30} \texttt{kg}$, space with $ M_\odot (\texttt{km}) =  1.477~\texttt{km}$, and time with $M_\odot (\texttt{s}) = 4.92674\times 10^{-6} \texttt{s}$. 

\section{Theoretical Framework: the Weak-Field Region}
\label{sec:weakfield}
\subsection{The Keplerian Problem}
\label{ssec:kepler}

Let us start with an eccentric pair of two black holes reduced to material points of masses $m_1$ and $m_2$, separated by a distance $r$, each revolving in elliptical orbits around a common center of mass. 
This model is known as the \emph{two-body problem}.
Imposing the conservation of momentum, this is further simplified to the \emph{one-body problem} for a single particle of \emph{reduced mass} $\mu=m_1m_2/M$ and position $r$, moving in the external gravitational field created by the mass $M=m_1+m_2$ located at the center of mass. 
Due to the conservation of angular momentum, the position vector will be confined to the plane, allowing us to simplify the dynamics to a two-dimensional problem.
Moreover, because $r=r(\cos \theta, \sin \theta)$, we only need to know how the angle $\theta$ varies in time, to describe the motion of the point mass $\mu$ \cite{fitzpatrick:book}.
This deceptively simple, now \emph{one-dimensional} ansatz is also known as the \emph{Keplerina problem}, but even in Newtonian gravity it is very complicated to solve. 
At the heart of this model lies Kepler’s equation, with its still unknown exact analytical solution, although scientists tried to find one for centuries.
Luckily, we have analytical methods that approximate this solution with the desired degree of accuracy \cite{colwell:book}.

\begin{figure}[!ht]
\centering
\includegraphics[scale=0.75]{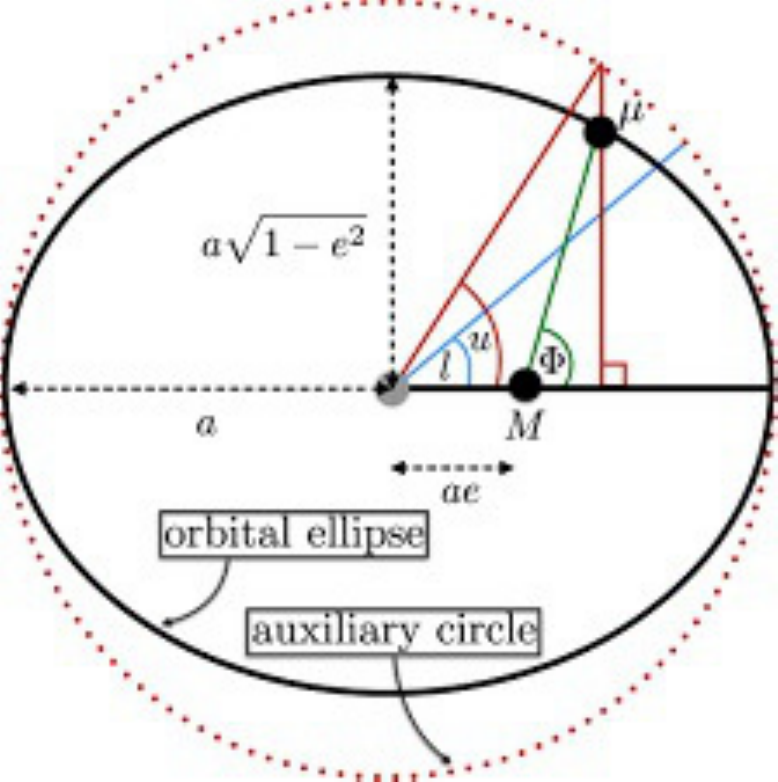}
\caption{Representation of the elliptical orbit as depicted in \cite{arXiv:1505.06208}}
\label{fig:orbit}
\end{figure}

At a a closer look, the elliptical orbit can be described by only four quantities: the two extreme points that bound the trajectory, $r_{-}$ (periastron, or point of closest approach to the origin) and $r_{+}$ (apoastron or maximum distance from the origin), the semi-major axis $a = \frac{1}{2}(r_{+} + r_{-})$, and the eccentricity 
\begin{equation}
e = \frac{r_+ - r_-}{r_+ + r_-}.
\end{equation}
We align the $x$-coordinate with the major axis of the ellipse and draw a circle of radius $a$ with origin in the center of the ellipse. 
Then, we raise a perpendicular from the x-axis through the orbiting point-mass $\mu$, thus projecting its motion onto this circle.
The motion is now reduced to a circular orbit on this \emph{reference circle} described by the angular coordinate $u$, called \emph{eccentric anomaly} (see Fig. \ref{fig:orbit}). 

In order to calculate the angle $u$, we must use the Kepler's equation, deduced from purely geometric considerations as a relationship between this eccentric anomaly and the eccentricity of the ellipse:
\begin{equation}
\ell = u - e \sin u,
\label{eq:kepler}
\end{equation}
where $\ell =w(t-t_0) $ is called the  \emph{mean anomaly}, and $w = 2 \pi t/P$ is the \emph{mean motion}, namely the angular speed of the body $\mu$ averaged over the period $P$ of the orbit. 
This quantity is measured with respect to the periastron, between $\ell = 0$ and $\ell = 2 \pi$ and given by Kepler's third law, $w=\sqrt{M a^{-3} }$ \cite{arXiv:1609.00915}.
Once we know this angle, we calculate the separation with the \emph{shape equation} $r = a (1 - e \cos u)$. The location of each star in the binary with respect to the center of mass will be recovered with $|r_{1,2}|=r {m_{1,2}}/M$.
Kepler's equation \eref{eq:kepler} is transcendental because it contains a trigonometric function  and cannot be computed exactly, although an exact solution should exist. 

Over time, many numerical and analytical approximate solutions have been proposed, and with the aid of computers we can calculate them with high degree of accuracy \cite{colwell:book}.
We will use Bessel's approach who, using the well known property that any continuous function can be expanded in Fourier series, wrote the solution to Kepler's equation \eref{eq:kepler} as:
\begin{eqnarray}
u = \ell + \sum_{i=1}^{\infty} a_i \sin (i \ell), ~~
a_i = \frac2 i J_i(i e) \sin (i \ell),
  \label{eq:kfourier}
\end{eqnarray}
where $J_i(x)$ are the modified Bessel function of the first kind \cite{arfken:book}.
This gives an analytical expression for the eccentric anomaly, which is geometrically related to the orbital phase $\phi$, also known as the \emph{true anomaly}, by the equation:
\begin{equation}
{\phi}= 2 \tan^{-1} \left(\sqrt{\frac{1+e}{1-e}} \tan \frac{u}{2} \right).
\label{eq:phi}
\end{equation}
Once we know the orbital phase, in Newtonian gravity the problem is solved.

\subsection{The quasi-Keplerian Problem}
\label{ssec:qkepler}

Up to this point we have not taken into account the loss of orbital energy due to the emission of GWs.
Let us now include this essential component into our treatment of the binary's orbit, while assuming that the orbit decays much slower compared to the orbital period. 
This important conjecture, called the \emph{adiabatic approximation}, allows us to model the emission of GW with the \emph{balance equation}, stating that the loss of orbital energy is balanced by the flux of GWs for a binary system in a slowly-evolving circular orbit:
\begin{equation}
-\frac{dE(t)}{dt}= \dot E(t) ={\cal F}(t).
\label{eq:balPN}
\end{equation}
Here, $E(t)$ is the binding energy of the binary and ${\cal F}(t)$ is the energy flux emitted in GWs.
This equation is valid for most of the inspiral, as long as the separation $r$ is large and the orbital speed is much smaller than the speed of light, such that $x_{PN}\ll1$. 
In this region, the gravitational field is weak and the GR equations can be solved in terms of power series expansions of the parameter $x_{PN}$, starting from the familiar Newtonian two-body equation of motion. 
This is the post-Newtonian formalism, introduced in fact by Einstein \cite{Einstein:PN}, and to the reader interested in learning about it we recommend \cite{book:gravity, arXiv:1310.1528}. 
This powerful and accurate method provides the expressions for the binding energy and the emitted flux, enabling us to integrate the balance equation \eref{eq:balPN} and to obtain the evolution equation for the parameter $x_{PN}$
\begin{equation}
\frac{d x_{PN}(t)}{dt} = \dot x_{PN}(t) = -\frac{{\cal F}(t)}{(dE(t)/dx_{PN})}. 
\label{eq:xPN}
\end{equation}

The dynamics of the binary becomes much more complicated when we add eccentricity to the orbit. 
Due to the ellipticity, the curvature of the orbit changes, making the angle of the tangent to the orbit vary along the trajectory, effect described by the \emph{extrinsic curvature} in GR. 
This theory assumes a four-dimensional spacetime, where the orbit depends on how we chose the three-dimensional spatial slice in this 4D geometry.
This means that how we track the evolution of the eccentricity is not invariant of the coordinate system, and will depend on how we define it. In mathematical language, the eccentricity is not \emph{gauge invariant} and depends on our choice of coordinate system. 
To circumvent this problem, we employ again the adiabatic approximation and assume that the change in eccentricity due to the emission of GWs occurs on a timescale much longer than the orbital timescale.
This allows us to average the eccentricity over the orbital period and to track its variation only from one orbit to the next, computing thus the orbit-averaged version of the evolution equation for the eccentricity.
As a note of caution, this approximation breaks down in the late inspiral of the binary, when the orbit becomes quasi-circular and the orbital velocity approaches the speed of light. 
In this region, starting roughly around $v \approx 0.25$, the emission of GWs forces the orbit to decay faster and to circularize under the effect of radiation-reaction \cite{arXiv:1801.09009}.
As long as the inspiral does not reach this region, the orbit decays slowly enough for the motion within one orbit to be conservative, keeping the adiabatic approximation valid with high degree of accuracy. 

What happens thought when the eccentricity decreases in time due to the emission of GWs, while the orbital velocity increases? 
This is the two-body problem in GR, which does not have an analytic solution.
Nevertheless, we can generalize the solution to Kepler's equation to include these relativistic effects, following \cite{arXiv:1707.02088}. 
\begin{eqnarray}
u &= \ell + 2\sum_{i=1}^{\infty} A_i \sin (i \ell), 
\label{eq:kPN} \\ 
A_i &= \frac1i J_i(i e_t) + \sum_{j=1}^{\infty}\alpha_j [J_{i+j}(i e_t) - J_{i-j}(i e_t) ],
\label{eq:kAPN}
\end{eqnarray}
where $e_t$ is the time-dependent eccentricity, $J_i$ are again the modified Bessel functions, and $\alpha_j$ are the perturbative, post-Newtonian corrections, that depend on $e_t$, on the parameter $x_{PN}$, as well as on the chosen coordinate system.
This slightly perturbed form of Kepler's equation \eref{eq:kPN} can still accurately describe the orbital evolution in GR. 
This formalism is called the \emph{quasi-Keplerian} parametrization and comes at a price, namely we must take into account three eccentricity parameters to include the three important changes in the orbit introduced in GR, namely: (1) $e_t$ for the change in the orbital period due to the emission of GW, (2) $e_{\theta}$ for the relativistic precession of periastron and (3) $e_r$ for the shrinking of the orbit  \cite{arXiv:gr-qc/0404128, arXiv:gr-qc/0407049, arXiv:gr-qc/0603056}. 
These eccentricities depend all on the energy and angular momentum of the binary and thus on each other, allowing us to choose in our calculation only one of them, namely $e_t$, as is customary in the literature \cite{arXiv:1810.03521}. 
Finding analytic solutions to the quasi-Keplerian equation is, as expected, much more complicated now, but it has been achieved up to 3PN order in accuracy \cite{arXiv:1707.02088, arXiv:1801.08542, arXiv:1903.05203, arXiv:1904.11814}.

\subsection{The Inspiral Waveform}
\label{inspiral}
Up to this point, we described in detail the motion of the binary in the weak-field region. 
Armed with this knowledge, we can turn now to our main objective, namely to calculate  the analytical form of the emitted GW.
We will start with the formula for the dimensionless strain, 
\begin{equation}
h(t) =  h_{+}(t) - \mathrm{i} h_{\times}(t)=A(t) e^{-\mathrm{i} \phi_{GW}(t)}, 
\label{eq:h+x}
\end{equation}
where the two strain components ($h_{+}(t)$, $h_{\times}(t)$) describe the two possible polarizations of the wave, $\phi_{GW}(t)$ is the phase of the GW and $A(t)$ is its complex amplitude.
At the \emph{optimal orientation}, when the spherical polar angles of the observer are $\theta = \varphi = 0$ \cite{arXiv:0806.1037}, the binary is face-on and right overhead the detector, at some distance $R$ from it. 
In this case, the amplitude of GW is maximum, and is given by \cite{arXiv:1709.02007}:
\begin{eqnarray}
\label{eq:Ains}
A(t) &= A_{\cal R}(t) + \mathrm{i} A_{\cal I}(t) \\
\nonumber 
&= -\frac{2 \mu }{R} \left ( \dot r^2(t) + r^2(t) \dot \phi^2(t) + \frac{M}{r(t)} + 2 \mathrm{i}r(t)  \dot r(t) \dot \phi(t) \right ).
\end{eqnarray} 

GWs are \emph{quadrupolar} due to the law of conservation of linear and angular momentum, such that the signal goes through  maxima and minima twice during one orbital cycle, thus: $\phi_{GW}(t) = 2 \phi(t)$.
The strain from eq.\eref{eq:h+x} is usually decomposed in spin-weighted ($s=-2$) spherical harmonics modes $(l, m)$ \cite{harmsylm}, 
\begin{equation}
h(t)=\sum_{l=2}^{\infty}\sum_{m=-l}^{l} {}_{-2} Y_{lm} h_{lm}(t).
\end{equation}
The leading mode $(l=2,m=\pm 2)$ typically dominates the sum ($h(t) \approx \sqrt{5/4\pi} h_{22}(t)$). 

To calculate the strain in eq.\eref{eq:h+x} we need to know the equations of motion for the orbital phase $\phi(t)$ and the orbital separation $r(t)$.
In the PN approximation those quantities are written as powers of the small parameter $x_{PN}(t)$, as shown below:
\begin{equation}
r_{PN}(t) = \frac{M}{x_{PN}(t)} \sum_{j=0}^{N} \rho_{j PN} x_{PN}^{j}(t) ,
\label{eq:rPN}
\end{equation}
\begin{equation}
\dot \phi_{PN}(t) = \frac{x_{PN}^{3/2}(t)}{M} \sum_{j=0}^{N} \varphi _{jPN} x_{PN}^{j}(t) .
\label{eq:phiPN}
\end{equation}
Here $N$ is the order of the post-Newtonian expansion and the terms ($\rho_{jPN},  \varphi _{jPN}$) are numerical expansion coefficients. 
To calculate those expressions we need to know how $x_{PN}(t)$ evolves in time, as given by the balance equation \eref{eq:xPN}. 
There are several ways of further expanding this equation in post-Newtonian powers of $x_{PN}(t)$, and we will choose the \emph{TaylorT4 approximant} \cite{arXiv:gr-qc/0211087}, which gives for $x_{PN}(t)$ the equation \cite{arXiv:1901.07038}: 
\begin{equation}
\dot x_{PN}(t)=\frac{x_{PN}^{5}(t)}{M} \sum_{j=0}^{N} \xi_{jPN} x_{PN}^{j}(t) + \dot x_{HT}(t).
\label{eq:dxPN}
\end{equation}
Note that $\dot x_{HT}$(t) are called \textit{hereditary} terms and enter in the equation as fractional $j/2$  PN corrections \cite{arXiv:1609.05933}. 
Those terms describe the nonlinear interaction of the GWs propagating away from the sources at later time with the GWs that propagated towards the source at an earlier time (\emph{memory effects}) and again back-scattered by the spacetime curvature of the binary (\emph{tail effects}). 
Those hereditary terms, containing both memory and tail effects, depend on all the past dynamical history of the source (see \cite{arXiv:1310.1528, arXiv:1408.5505} for details). 

We must add to eq.\eref{eq:dxPN} the following equation of motion for the eccentricity, because the eccentricity diminishes as the orbit circularizes while the binary inspirals emitting GWs as it approaches the merger, 
\begin{equation}
\dot e_{PN}(t)=\frac{x_{PN}^4(t)}{M} \sum_{j=0}^{N} \epsilon_{jPN} x_{PN}^{j}(t) + \dot e_{HT}(t).
\label{eq:dePN}
\end{equation}

Next, we find the evolution of the mean anomaly $\ell$ that describes how the orbital angular velocity changes in the course of an orbital evolution due to the eccentricity. 
\begin{equation}
\dot \ell_{PN}(t) = w(t) = \frac{x_{PN}^{3/2}(t)}{M} \sum_{j=0}^{N} \lambda _{jPN} x_{PN}^{j/2}(t). 
\label{eq:dlPN}
\end{equation}
After we gather the expansion coefficients $(\xi_{jPN},  \epsilon_{jPN}, \lambda _{jPN})$ and calculate how the key quantities $e_{PN}(t)$,  $x_{PN}(t)$ and $l_{PN}(t)$ evolve in time with eqs.\eref{eq:dxPN}, \eref{eq:dePN} and \eref{eq:dlPN}, we must solve the quasi-Keplerian equation \eref{eq:kPN} to find the eccentric anomaly $u_{PN}$.
The last two steps, before obtaining the strain in eq.\eref{eq:h+x}, are to integrate equation \eref{eq:phiPN} to find the orbital phase $\phi(t)$, and to calculate the separation $r(t)$ with equation \eref{eq:rPN}.

\section{Theoretical Framework: the Strong-Field Region}
\label{sec:strongfield}
\subsection{Location, Location, Location}
\label{locations}
Let us now attempt to tackle the problem of the binary motion when the orbital velocity approaches the speed of light and the PN formalism can no longer be applied.

Going beyond this approximation brings us up against the strong gravitational field region, which can be described accurately only by Einstein equations of GR that connect the gravitational field with the geometry of spacetime through the curvature. 

The transition between the weak and strong field starts at the last stable orbit a particle would have when moving around the central black hole, called the \emph{innermost stable circular orbit} (ISCO). 
This is well defined only for a small point-like body orbiting a much more massive one, and becomes increasingly less delimited as the masses of the two bodies are comparable \cite{arXiv:gr-qc/0703053}. 
We will stay within the one-body problem model considered before, in which a particle of mass $\mu$ is orbits a central black hole of mass $M$, only this time we will extend the size of the central body from a material point to a static Schwarzschild black hole of radius $r_{Sch}$ (the Schwarzschild radius). 
In this case, it is analytically proved that ISCO is located at $r_{ISCO} = 6 M$ or $3 r_{Sch}$ in geometrical units \cite{arXiv:gr-qc/0507692}. 
If the central black hole is spinning though, ISCO will move closer or further away from this position, depending on the direction of the particle's orbit relative to the black hole's spin $S_i$. 
The analytical expression for the dimensionless radius of the last stable prograde orbit is given below \cite{bardeen, arXiv:gr-qc/0003032}. 
\begin{equation}
{\tilde r_{ISCO}} = \frac{r_{ISCO}}{M}  = 3 + Z_2 - \sqrt{(3 - Z_1)(3 + Z_1 + 2 Z_2)}.
\label{eq:rISCO}
\end{equation}
Here $ Z_1= 1 + (1 - \chi_f^2)^{1/3}[(1 + \chi_f)^{1/3}+ (1 - \chi_f)^{1/3}]$ and $Z_2 = \sqrt{3 \chi_f^2 + Z_1^2}$ with $\chi_f=S_f/M_f^2$ is the dimensionless spin of the final black hole.

Beyond ISCO we enter the \emph{plunge} region, which ends at the \emph{light ring} (LR) --  the last stable photon orbit. 
The LR is located at $r_{LR} = 1.5 r_{Sch} = 3 M$ and changes if the black hole is spinning, depending on the orbit of the photon with respect to its spin. 
Its location, calculated analytically in \cite{bardeen, arXiv:gr-qc/0001013} for a rotating black hole, plays an important role in the matching procedure, and for the prograde orbit is given by: 
\begin{equation}
r_{LR} = 2 M \left [1 + \cos{\left(\frac{2}{3}\cos^{-1}(-\chi_{f})\right)}\right ].
\label{eq:rLR}
\end{equation}
After the LR, the \emph{coalescence} (or merger) begins, during which the two black holes come in contact and collide, forming a highly distorted common envelope, called \emph{apparent horizon} (AH) \cite{arXiv:gr-qc/9905039, arXiv:1511.07775}. 
This is a third dimensional surface, locally defined as the place where the photons directed outward from the interior region cannot escape to the exterior. 
The AH coincides with the boundary of the black hole, namely the \emph{event horizon} (EH), only for static black holes $r_{EH} = 2 M$. This is because the EH is a four dimensional surface in GR and it's clear location can be calculated only at the end of all times.
In dynamical spacetimes the EH is approximated with the AH, which is used to mark the surface of the distorted black hole formed by collision \cite{ arXiv:1703.00118}. 
The general expression for the radius of the outer event horizon of a rotating (Kerr) black hole is:
\begin{equation}
r_{AH} = r_{+} = M(1 + \sqrt{1 - \chi_f^2}).
\label{eq:rAH}
\end{equation}
After the merger, we must switch models from the one-body problem, to the \emph{close limit approximation} (CLA) \cite{gr-qc/9402039}. This is an analytical approach that uses the perturbation theory to analyze how a highly perturbed Schwarzschild black hole rings down the remaining energy in form of GWs and becomes quiescent \cite{arXiv:gr-qc/9909021, arXiv:0910.4593}. 
The amplitude of the emitted GWs increases between the ISCO and the LR, reaches its peak (or chirps) around the AH, and diminishes in the ringdown phase, while the frequency increases monotonically.

We separate thus the strong field in three regions: plunge, merger and ringdown, delimited by three locations: $r_{ISCO}$ \eref{eq:rISCO}, $r_{LR}$ \eref{eq:rLR}, and $r_{AH}$ \eref{eq:rLR}.
These locations enter in the analytical calculation of the GWs emitted in the strong field region and play a key role in determining the optimal place where we should glue together the weak and strong field waveforms.

As a side note, it is worth mentioning that several authors \cite{arXiv:gr-qc/0003032, arXiv:1607.05661, arXiv:1611.00332} replace the location of the LR with the position of the \emph{minimal energy circular orbit} (MECO). 
This is based on the observation that no constraining connection exists between the LR, which marks a peak in the effective radial potential, and the dynamics of the spacetime during the collision, which is governed by the curvature potential, as is argued in \cite{arXiv:1609.00083}. 
However, in the same paper it is pointed out that those two locations are close enough to each other that the wavelength of the GW is insensitive to such small variation in the position of the peak. 
We will consider in this work that the LR is an accurate approximation for the peak of the curvature potential. 

\subsection{The Merger Waveform}
\label{ssec:BoB}
For the merger, we employ the \emph{Backward-one-Body} (BoB) model, introduced in \cite{arXiv:1810.00040}, that calculates the 
strain of the GWs using the formula:
\begin{equation}
h_{BoB}(t) = -\frac{A_{BoB}(t)}{\Omega^2(t)}e^{-i\phi_{BoB}(t)}
\label{eq:strainBoB}
\end{equation}
where the amplitude has the form:
\begin{equation}
A(t)_{BoB} = A_0 \mbox{sech} \left(\frac{t-t_0}{\tau_{QNM}}\right)
\label{eq:ampBoB}
\end{equation}
Here $A_0$ is an integration constant that enters as a scaling factor, $t_0$ is the time at which the strain of the GW reaches its peak amplitude, and $\tau_{QNM}$ is the damping time necessary for the final black hole to settle down after the collision.
During this time, the highly perturbed remnant object releases its energy in GWs as characteristic exponentially damped \emph{quasi-normal} modes (QNM).
We can find the damping time using the definition \cite{arXiv:gr-qc/0512160}:
\begin{equation}
\tau_{QNM} = 2 \frac{Q_{QNM}}{\Omega_{QNM}}.
\label{eq:tau}
\end{equation}
where the quality factor $Q_{QNM}$ is a dimensionless quantity measuring the number of oscillations observed before a mode gets attenuated by a factor of $e^{2 \pi}$ \cite{arXiv:0903.0338}, and  $\Omega_{QNM}$ is the QNM frequency.
Whatever initial data we provide inside the $r < 3M$ region where the peak of the potential barrier is located, this barrier will filter it, and an outside observer will detect only the QNM ringing.
We employ the following polynomial fit for the pair ($\Omega_{QNM}, Q_{QNM}$):
\begin{equation}
\label{eq:OQNM}
M_f \Omega_{QNM} = f_1 + f_2(1 - \chi_{f})^{f_3},
\end{equation}
\begin{equation}
\label{eq:QQNM}
Q_{QNM} = q_1 + q_2(1 - \chi_{f})^{q_3} 
\end{equation}
with the set of coefficients $f_{1,2,3}$ and $q_{1,2,3}$ given in \cite{arXiv:0905.2975}. 

The phase in eq.\eref{eq:strainBoB} is obtained by integrating the frequency $\Omega(t)$:
\begin{equation}
\Phi_{BoB}(t) = \int \Omega_{BoB}(t) dt,
\label{eq:phiBoB} 
\end{equation}
which is given in this model by the equation:
\begin{equation}
\Omega_{BoB}(t) = \left( \Omega_i^4+\kappa \left[ \tanh \left(\frac{t-t_0}{\tau}\right) - \tanh \left(\frac{t_i-t_0}{\tau}\right) \right]\right)^{1/4}
\label{eq:omgBoB}
\end{equation}
This equation makes use of the parameter $k$ given by the expression:
\begin{equation}
\kappa = \left[ \frac{\Omega_{QNM}^4-\Omega_i^4}{1 -\tanh\left(\frac{t_i-t_0}{\tau} \right)}\right]
\label{eq:constk}
\end{equation}
The initial frequency $\Omega_i$ corresponds to the frequency at the end of the inspiral, around the LR, where the PN approximation breaks down.
The most important quantity is the initial time $t_i$, which is analytically obtained by replacing eq.\eref{eq:constk} in eq.\eref{eq:omgBoB}:
\begin{equation}
t_i = t_0 - \frac{\tau_{QNM}}{2}\ln \left( \frac{ \Omega_{QNM}^4 - \Omega_i^4}{2\tau\Omega_i^3\dot{\Omega}_i} - 1 \right). 
\label{eq:tisimp}
\end{equation}
This equation is essential, being the key ingredient that determines the exact time when the merger waveform is added to the inspiral when constructing the total strain.

\section{Results: Building the Compete Waveform}
\label{sec:complete}

\subsection{It's about time}
\label{ssec:time}
Before we turn all this formalism into concrete waveforms, let us take a better look at the independent variable dictating the evolution of the whole system, namely the time coordinate, which is taken to coincide with the proper time of distant observers.

Let's start with the expression given in \cite{Peters:1964} for the \emph{coalescence time} until a binary in quasi-circular orbit reaches the merger, from a known initial separation or semi-major axis $r_0$, due to the emission of GW. In geometrical units, this formula is
\begin{equation}
T_c = \frac{5}{256}\frac{r^4_0}{m_1 m_2 (m_1+m_2)}
= \frac{5}{256}\frac{r^4_0 (1+q)^2}{M^3 q}.
\label{eq:tcoal}
\end{equation} 
where $q = m_1/m_2$ is the mass ratio of the binary. The addition of a small initial eccentricity $e_0$ is accounted for by replacing $r_0$ in eq.\eref{eq:tcoal} with
\begin{equation}
a_0 =  r_0 \frac{8.70127(1 - e_0^2)}{(304 + 121 e_0^2)^{870/229}}.
\end{equation} 
Recently, improved estimations of this time, including $1^{st}$ order PN perturbations, were given in 
 \cite{arXiv:1905.06086, arXiv:1911.06024}. 
The choice of the initial separation $a_0$ is limited by the low frequency cutoff of Earth's seismic activity $f_{low}$, which dictates the threshold value of GWs detection band for aLIGO. For a cutoff frequency $f_{low} = 10~\texttt{Hz}$, we obtain from Kepler's third law $ a_0=( {M}/{\pi^{2} f_{low}^{2}})^{1/3} $.
For an equal mass binary of total mass $M = 50M_{\odot}$, this amounts to a separation of about $1278 M_{\odot} = 1888~\texttt{km}$, and gives a coalescence time $T_c = 8.22~\texttt{s}$ of the GW signal in the aLIGO band. 
The coalescence time decreases for eccentric binaries, thus if we assume for example that the binary enters the aLIGO band with $e_0 = 0.25$, it will take $T_c = 6.11 ~\texttt{s}$ to collide, which will be enough time for the eccentricity to be detected.  
 The LISA band cutoff frequency however is much lower, $f_{low,LISA} = 10^{-4}~\texttt{Hz}$, which, for a $50M_{\odot}$ equal-mass binary gives a separation of $4 \times 10^6~\texttt{km}$, and consequently a collision time of about five millions years.
 Thus systems with large eccentricities are expected to be regularly detected by space-borne instruments. Conversely, GW signals lasting only a few seconds in the LISA band will come from billions solar masses binaries.  

The issue we must tackle next is to find an appropriate estimate for the time when the inspiral model breaks down and must be replaced with the merger model.
As we mentioned before, the transition between the weak and strong regime happens around the last stable orbit (ISCO), located around $r_{ISCO}=6M$. 
The problem with the start of the merger phase at ISCO is, as we explained previously in Sec.\ref{locations}, that ISCO is not well defined for comparable mass binaries. 
Fortunately, we can still define an approximate ISCO even in this case, closer to the light-ring, given in eq.\eref{eq:rLR}, which in the non-spinning case ($\chi_f = 0$) is $r_{LR} = 4M$ \cite{arXiv:0706.3732}. 
Using eq.\eref{eq:tcoal} for $r_0 = r_{LR} = 4 M$, we get $t = 20 M$, which corresponds to the initial time $t_i$ for the strong region. 
Here, in the $0^{th}$ PN approximation, the orbital frequency will be $M \omega_{LR} = 0.125$, which represents a lower limit for the orbital frequency of the inspiral model predicted by the PN theory around $t_{LR}$.
Another constraint, offered by eq.\eref{eq:tisimp}, provides a third independent check for the relationship between this time and the frequency at which the merger waveform joins smoothly with the inspiral. 
The reader is warned that some arbitrariness will be still inherent due to the assumptions we will make, because there is no unique way of defining this transition time unambiguously.

Lastly, we will choose the time at which the amplitude of the fundamental mode has a maximum, as the origin of the time axis $t_{0}=0$. 

\subsection{Eccentric Slowness}
\label{ssec:eccentricity}
Our procedure for the implementation of the inspiral waveform follows the outline given in Sec.\ref{inspiral}.
First, we implement all the expressions for the coefficients that appear in equations \eref{eq:rPN}, \eref{eq:phiPN}, \eref{eq:dxPN}, \eref{eq:dePN} and \eref{eq:dlPN}. 
Those coefficients are currently known up to the $4^{th}$ leading PN order, with up to 6PN quasi-circular correction terms and hereditary corrections \cite{arXiv:1601.05588, arXiv:1707.09289, arXiv:1811.08966, arXiv:1902.09801}. Impressive efforts are continuously made to calculate new terms in the PN expansion \cite{arXiv:1903.05113, arXiv:2005.13367, arXiv:2003.11891, arXiv:2004.05407, arXiv:2101.08630}. 
Following \cite{arXiv:1609.05933}, we include up to 6PN self-force and hereditary correction terms, to increase the accuracy in modeling the region near the merger.  
We use \texttt{NDSolve} \cite{ndsolve} within the \texttt{Wolfram Mathematica} symbolic computation program \cite{Mathematica} to solve the coupled equations of motion \eref{eq:dxPN} and \eref{eq:dePN}, with the $\xi_{jPN}$ terms as in \cite{arXiv:1609.05933} and the eccentricity coefficients $\epsilon_{j_{PN}}$ from \cite{arXiv:0806.1037, arXiv:0908.3854}. 
We start from the aLIGO cutoff frequency $f_{low}$, work with $20$ precision digits and pick the collisions time $T_{s}$ when the equations become stiff and fail to yield a solution, then mark $t_{LR} = T_{s} - 20M$ as the end of the weak field region.

\begin{figure}[!ht]
\centering
\includegraphics[scale=0.75]{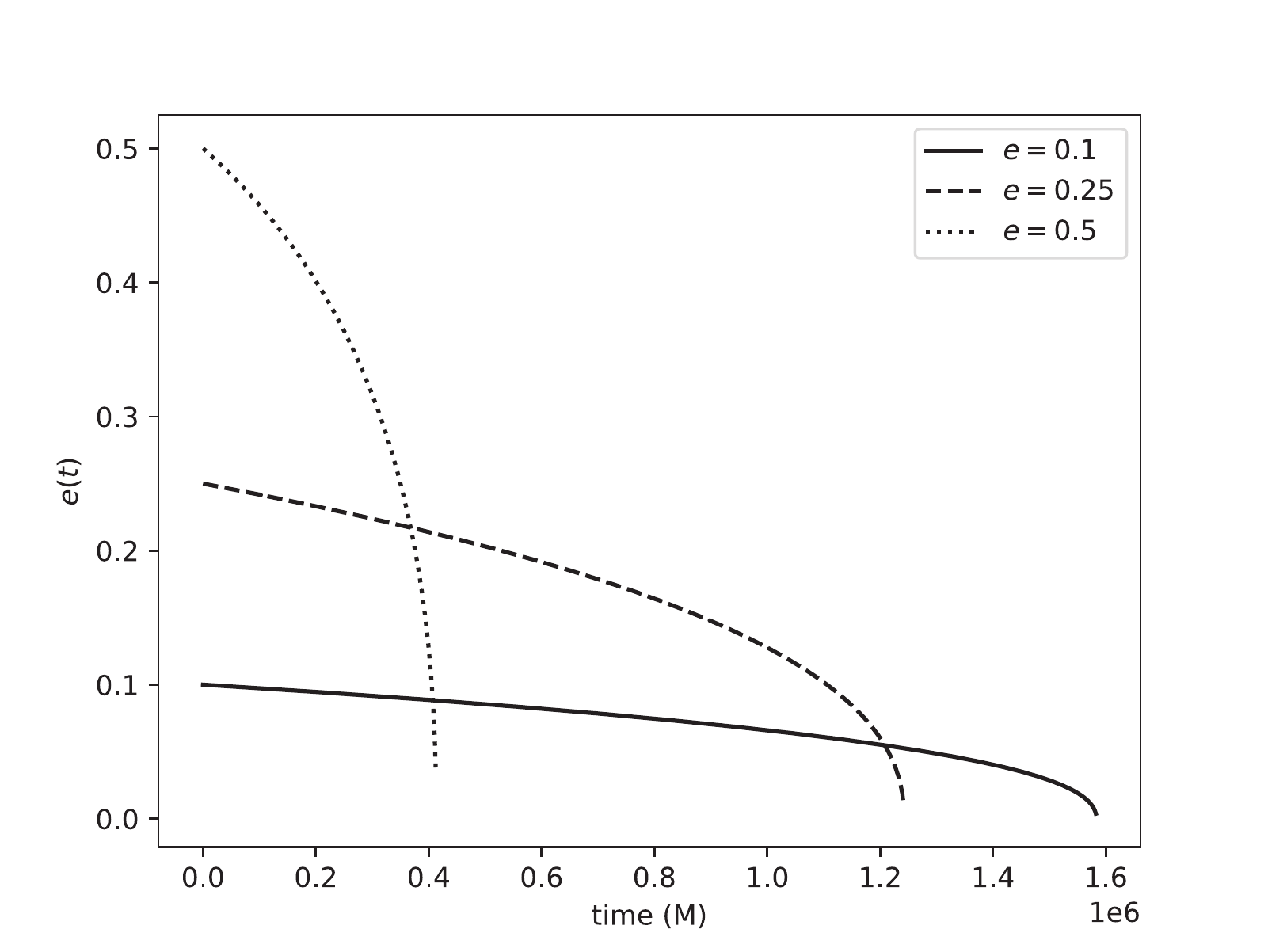}
\caption{Time evolution of the eccentricity, for an equal mass binary of $50 M_\odot$
(the time is measured in units of mass, $1M = 4.92674 \times 10^{-6} s$).}
\label{fig:q1ecc}
\end{figure}

Let's look now at the predicted evolution of the eccentricity, for an equal mass binary of $50 M_\odot$ total mass and three values for the eccentricity: $e_0=0.5, 0.25, 0.1$ at $f_{low}$. We choose the highest eccentricity to be $e_0=0.5$ because is close to the average eccentricity for wide binaries \cite{arXiv:2111.01789}. 
Fig.\ref{fig:q1ecc} shows that collisions of highly eccentric binaries happen faster, thus we expect a higher number of GW detections to come from initially highly eccentric systems. 

\begin{figure}[!ht]
\centering
\includegraphics[scale=0.75]{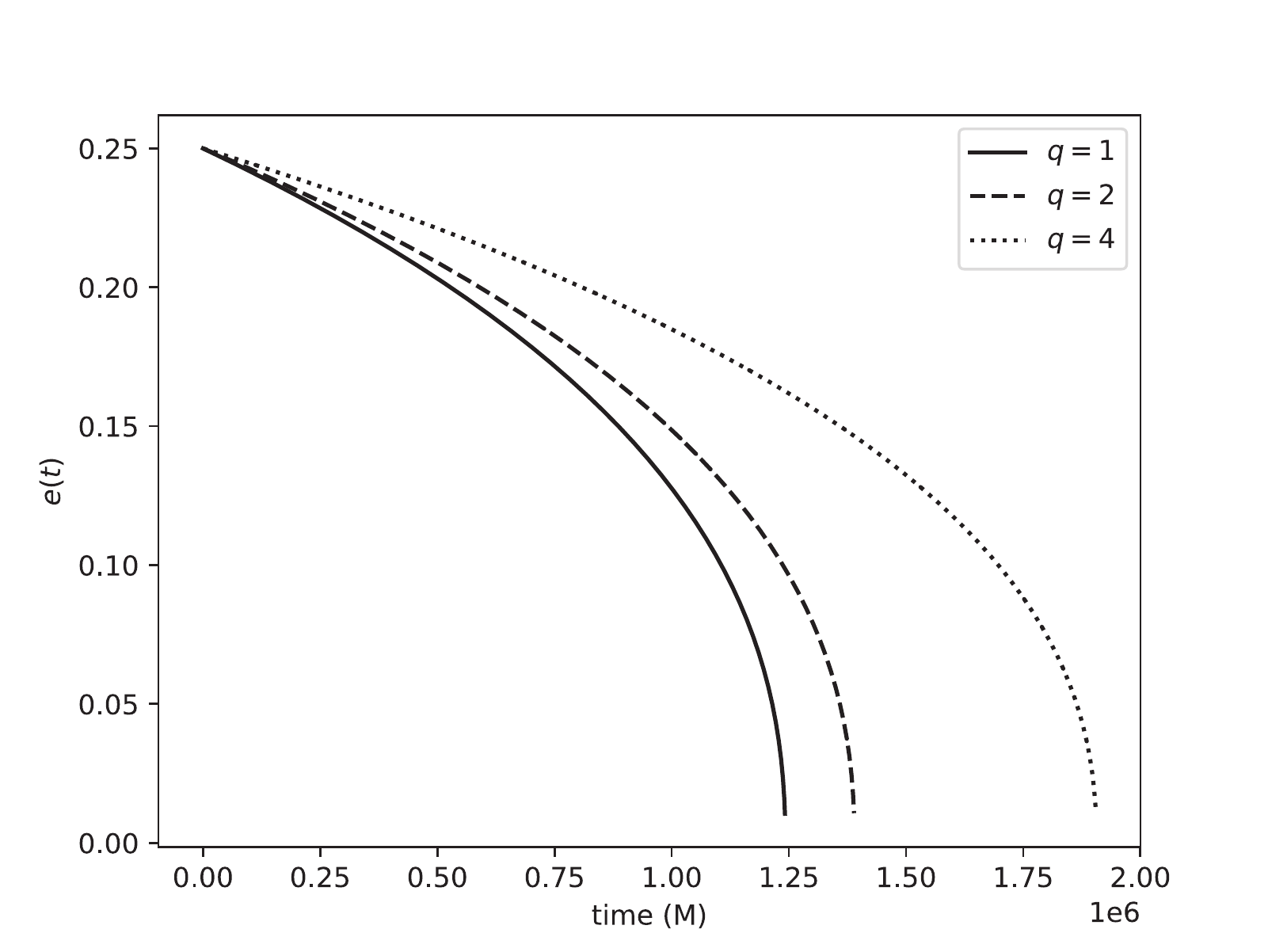}
\caption{Time evolution of the eccentricity, for a mass binary of $50 M_\odot$ and three mass ratios: $q=1,2,4$.}
\label{fig:qe025ecc}
\end{figure}

Fig. \ref{fig:qe025ecc}, shows the effect of the mass ratio on the eccentricity evolution, starting from $e_0 = 0.25$ eccentricity at cutoff frequency, for a $50 M_\odot$ mass binary with three mass ratios $q = 4, 2, 1$.
We see that small-mass ratio binaries take less time to merge than high-mass ratio ones, which has the expected consequence that more GW signals will come from equal or nearly equal mass binaries. 

\begin{figure}[!ht]
\centering
\includegraphics[scale=0.75]{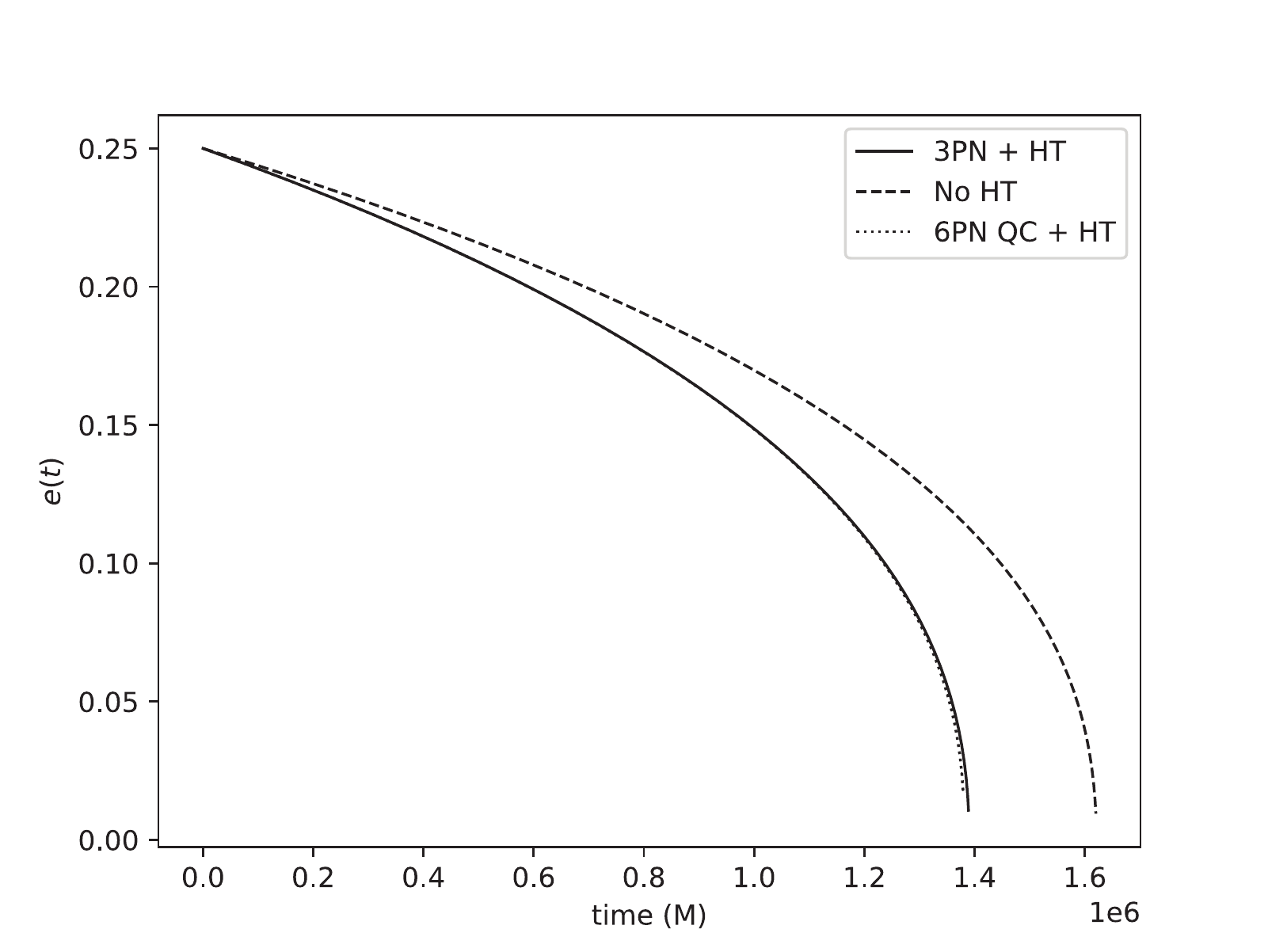}
\caption{Time evolution of the eccentricity, for a mass binary of $50 M_\odot$, mass ratio $q=2$ and eccentricity $e_0 = 0.25$, with three PN models.}
\label{fig:q2e025ecc}
\end{figure}

As expected, the evolution of the eccentricity depends also on the method used. 
Fig. \ref{fig:q2e025ecc} shows that taking into account the hereditary terms is important, because they add the fractional $n/2$ powers and cancel out terms depending on the chosen system of coordinates, thus ensuring the \emph{gauge-invariance} of the equations \cite{ arXiv:1609.05933}. 

In Table \ref{tab:eccl} we give the values for the eccentricity, slowness, and orbital frequency at the LR for an equal-mass binary with total mass $50$M, starting from  three different eccentricities, as well as a comparison between the numerical value of the collision time obtained when the equations become stiff $T_s$ and the collision time $T_c$ predicted by eq.\eref{eq:tcoal}.  
The slowness reaches a constant value at the LR ($x_{LR} \approx 0.178$), while the eccentricity decreases dramatically, more than $15$ times near LR, which proves that by the beginning of the strong field region the orbit of the system is circularized, even for large initial eccentricity, which can happen for capture binaries in dense clusters \cite{arXiv:1810.00901}.
It is reassuring that the eccentricity becomes close enough to zero by the time we reach the strong-field region, because the merger model is built for quasi-circular binaries. 
\small
\begin{table}[!htbp]
\caption{\label{tab:eccl}{Comparison between the computed time to LR and \\
the estimated collision time for three initial eccentricities at $f_{low}$, \\
$x_{PN}$ and the angular frequency at the LR for a $50 M_{\odot}$ binary.}}
\begin{indented}
\item[]\begin{tabular}{@{}llllll}
\br
$e_{0}$ &            $e_{LR}$         & $x_{LR}$ &   $M\Omega_{LR}$     & $T_{LR}  (s)$ & $T_c (s)$\\
\mr
$0.10$ & $5.952 \times 10^{-3}$ & $0.1782$ & $7.525\times 10^{-2}$ & $7.746 $ & $7.842$\\
\mr
$0.25$ & $1.671\times 10^{-2}$  & $0.1782$ & $7.526\times 10^{-2}$ & $6.074 $ & $6.113 $\\
\mr
$0.50$ & $5.779\times 10^{-2}$  & $0.1776$  & $7.483\times 10^{-2}$ & $ 2.010 $ & $2.251$\\
\br
\end{tabular}
\end{indented}
\label{tbl:LRvalues}
\end{table}
\normalsize

\subsection{All kinds of anomalies}
\label{ssec:anomaly}
In a previous work \cite{arXiv:1810.06160} we implemented an analytical calculation of GW from quasi-circular binaries up to the $3.5PN$ order, first by solving only eq.\eref{eq:dxPN}, then by relying on Kepler's third law to obtain the phase as: 
$\dot \phi_{PN}(t) = \omega_{PN}(t) = M^{-1} x_{PN}^{3/2}(t)$.
The addition of the eccentricity complicates things by introducing the angular anomaly which requires solving the transcendental Kepler's equation. 

The orbital angular velocity is now represented by the mean motion $w$, which changes in the course of an orbital evolution due to the presence of the eccentricity and is described by eq.\eref{eq:dlPN}.
To find the evolution of the mean anomaly $l(t)$ we integrate this equation with the $\lambda _{jPN}$ coefficients from \cite{arXiv:0806.1037}. 
With this quantity in hand, we proceed to calculating the eccentric anomaly given in eq.\eref{eq:kPN}, with the $3$PN coefficients $\alpha_j$ entering in the term $A_i$ taken from \cite{arXiv:1707.02088} for the Arnowitt-Deser-Misner (ADM) coordinates.
The evaluation of the Bessel functions is computationally expensive and in order to speed up the calculation we devised a method detailed in \cite{buskirk_2019}.
In here, we increase the truncation of the sum in the Bessel functions to $10$, tabulate $u(t)$, then apply a high order interpolating polynomial to keep the error around $10^{-20}$ precision.  

With the hardest part of the implementation out of our way, the next two steps are to calculate  the orbital separation $r(t)$ given by eq.\eref{eq:rPN} and the derivative of the phase as in eq.\eref{eq:phiPN}, with the coefficients $\rho_{jPN}$ and $\phi_{jPN}$ from \cite{arXiv:0806.1037}.
 In order to keep the desired precision and speed up the calculation, before we evaluate the time derivative of the separation, we use the same procedure of fitting a high-order polynomial to $r(t)$. 
Lastly, we solve for the phase $\phi(t)$ and build the strain with eq.\eref{eq:h+x}. 
We use the following formulas for the polarizations of the strain for optimal orientation of the observer \cite{arXiv:1810.06160}:
 \begin{eqnarray}
 h_{+}(t) = -2 \frac{M \eta}{R}\Bigg[ \Bigg. & \left( -\dot r(t) + r^2(t) \dot \phi^2(t) + \frac{M}{r(t)}\right )\cos 2 \phi(t)  \\
 \nonumber
               &+ 2 r(t) \dot r(t) \dot \phi(t) \sin 2 \phi (t) \Bigg. \Bigg]
\end{eqnarray}
 \begin{eqnarray}
 h_{\times}(t) = -2 \frac{M \eta}{R}\Bigg[ \Bigg. & \left( -\dot r(t) + r^2(t) \dot \phi^2(t) + \frac{M}{r(t)}\right )\sin 2 \phi(t)  \\
 \nonumber
               &- 2 r(t) \dot r(t) \dot \phi(t) \cos 2 \phi (t) \Bigg. \Bigg]
\end{eqnarray}
Those expressions can be extended to take into account the orientation of the binary with respect to the detector by performing three consecutive rotations from the frame of the orbit $\{x,y,z\}$ onto the frame of the observer $\{X,Y,Z\} = P \{x,y,z\}$, where $P$ is the rotation matrix written function of the three Euler angles, namely the longitude of the ascending node $\psi$, the inclination $\iota$ and the argument of the pericenter $\nu$, as described in \cite{arXiv:1609.00915}.  In this case, the true anomaly must be corrected to: $\phi \rightarrow \Phi = \phi - \nu$. We assume an optimal orientation of the detector and leave the inclusion of arbitrary orientation for a future work. 

We build our implementation hierarchically, constructing first the 2PN corrections, then adding the 3PN terms, and lastly the hereditary term $\dot x_{HT}$, while carefully testing our results against the zero eccentricity limit using the first GW detected signal GW150914 as a sanity check (as detailed in  \cite{buskirk_2019}).
The addition of the higher order hereditary corrections is necessary because, besides adding accuracy, it extends the validity of the inspiral model closer to the merger, increasing the overlap region with the merger model.

\begin{figure}[!ht]
\centering
\includegraphics[scale=0.75]{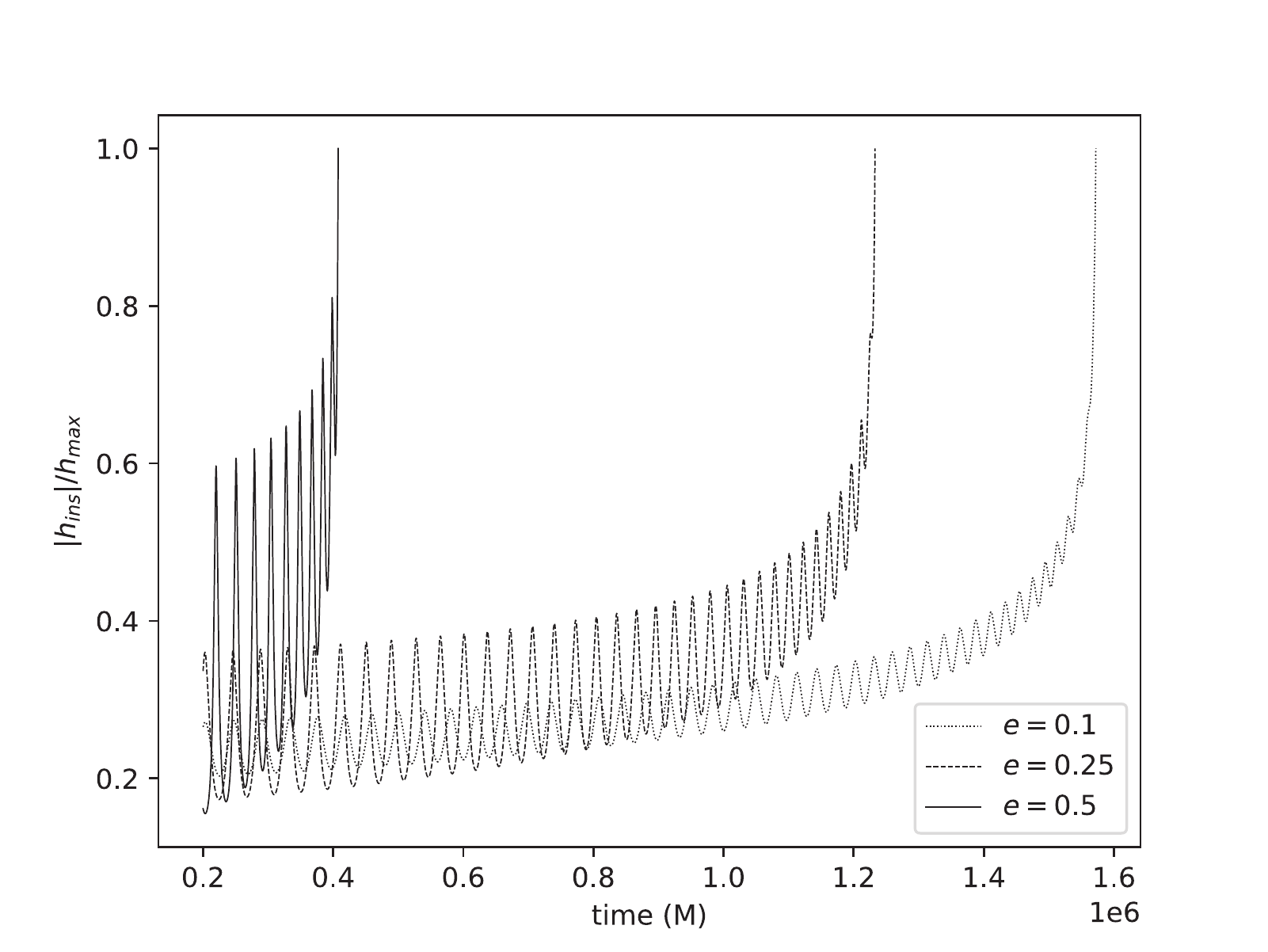}
\caption{Time evolution for the amplitude of the GW, with the time measured in units of mass.}
\label{fig:e01e025e05hamp}
\end{figure}

\begin{figure}[!ht]
\centering
\includegraphics[scale=0.75]{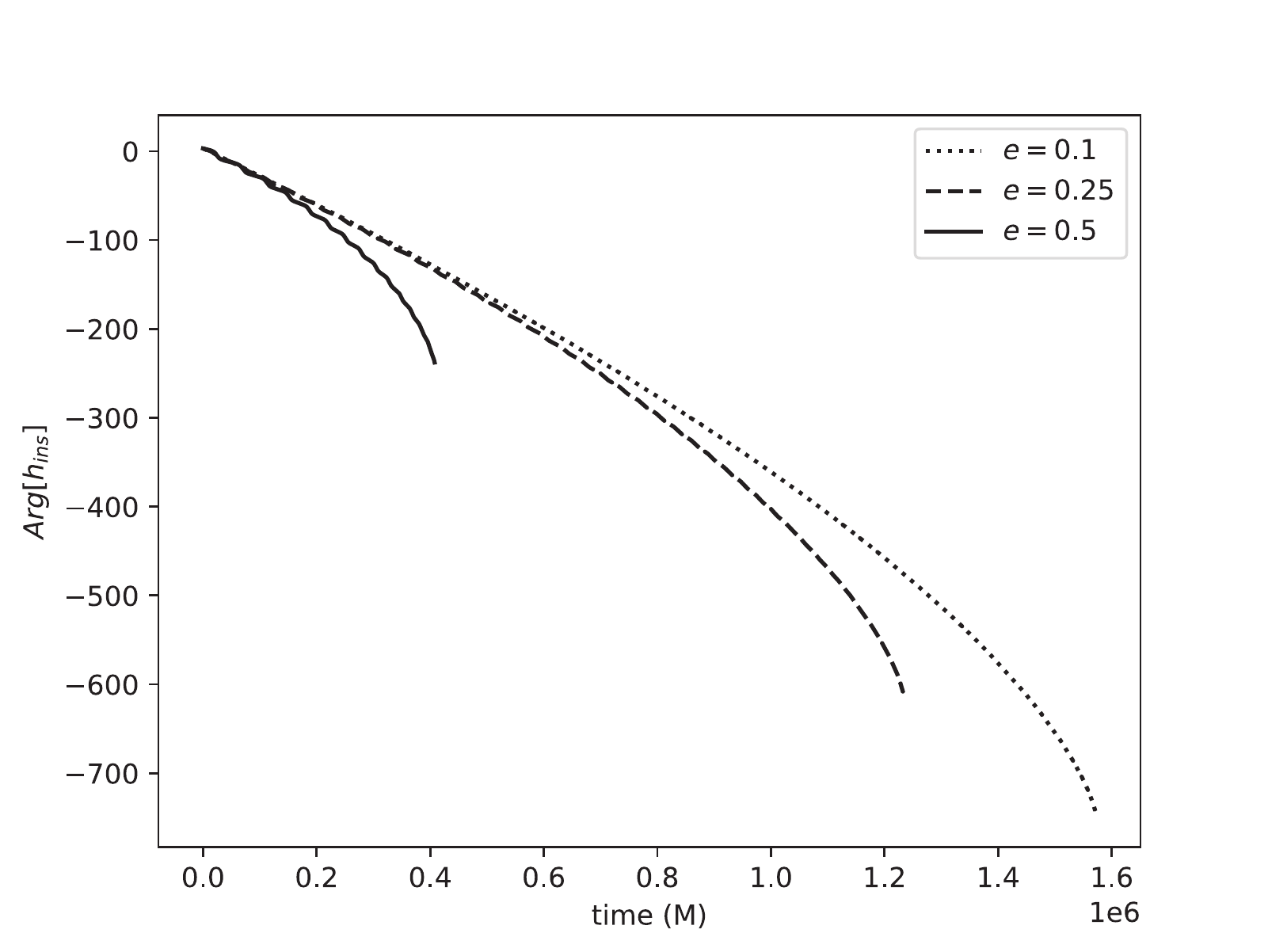}
\caption{Time evolution of the unwrapped argument of the strain, with time measured in units of mass.}
\label{fig:e01e025e05arg}
\end{figure}

We show in Figures \ref{fig:e01e025e05hamp} and \ref{fig:e01e025e05arg} the amplitude and the argument of the strain for optimal orientation of the observer. 
It is clear again, from Fig.\ref{fig:e01e025e05hamp}, that the eccentricity plays a vital role in determining the strength and duration of the signal, because highly eccentric system are more energetic but much shorter. 
In Fig.\ref{fig:e01e025e05arg} we plot the argument of the strain, which in absolute value is equal to twice the orbital phase  $\phi$.
We see that both the amplitude and the argument (phase) exhibit oscillations that diminish with the decrease of the eccentricity, until the orbit becomes quasi-circular near the merger.

\subsection{Prepare to Merge}
\label{ssec:before}
Before we proceed to model the merger part of the GW strain, we must know first the mass, spin and ringdown frequency of the final black hole, required by the BoB model as initial data. 
For the building of the inspiral strain we have not taken this spin into account because we treated the black holes as particles.
Once the separation becomes comparable to the LR radius, we cannot ignore the spin any longer.
If the spins are equal and antiparallel, or the black holes don't spin, then the final spin is $\chi_0 = 0.686$.
In this case eq.\eref{eq:rLR} shows that the location of the LR will move to about $r_{LR,\chi_0}=2.04M$, and the position of the apparent horizon, given by eq.\eref{eq:rAH}, will be around $r_{AH,\chi_0}=1.73M$.

With the final spin known, we proceed next to calculate the mass loss to GW energy, which subtracted from the total mass gives the mass of the final black hole.
 A simple estimate for the final mass is given by the Christodoulou formula \cite{christodoulou}
\begin{equation}
M_f= \sqrt{M_{irr}^2+\frac{S_f^2}{4 M_{irr}^2}} ,~\textrm{where}~M_{irr} = \sqrt{\frac{A_{AH}}{16 \pi}} = \frac{r_{AH}}{2}.
\end{equation}

The last essential ingredients we must add to the initial data for the merger model is the knowledge of fundamental QNM frequency $\Omega_{QNM}$ of the remnant black hole and the quality factor $Q_{QNM}$, which is necessary to calculate the damping time describing the decay in the amplitude of the GW.
For this we use eq.\eref{eq:OQNM}, \eref{eq:QQNM} with the set of coefficients for the dominant mode taken 
from \cite{echeverria}:
\begin{equation}
\label{eq:fiE}
f_1  = 1, ~f_2 = -0.63, ~f_3 = 0.3,~g_1 = 0,~ g_2 = 2, ~g_3 = -0.45.
\end{equation}
The damping time of the fundamental mode is given by eq.\eref{eq:tau}.

We calculate the final mass $M_f$, the dominant $(l=m=2)$ mode of the resonant frequency $\omega_{QNM}$, the corresponding quality factor $Q_{QNM}$ and the damping time $\tau$ for three non-spinning binary systems with mass rations $q =1, 2, 4$ and give the values in Table \ref{tab:mergeID}. We add to this table the PN values for $\Omega_i$ and its time derivative $\dot \Omega_i$, calculated in the inspiral model at the light ring.  
\small
\begin{table}[!htbp]
\caption{\label{tab:mergeID}{Initial data required for the BoB model.}}
\begin{indented}
\item[]\begin{tabular}{@{}llllllll}
\br
$q$ & $\chi_f$   & $M_f$      & $\omega_{QNM}$ & $Q_{QNM}$ & $\tau$          & $\Omega_i$ & $\dot \Omega_i$\\
\mr
$1$ & $0.6865$ & $0.9516$ & $0.5698$               & $3.301$      & $11.586$   &$7.529\times10^{-2}$ & $8.577\times10^{-4}$\\
\mr
$2$ & $0.6231$  & $0.9612$ & $0.5385$              & $3.056$      & $11.349 $   &$7.414\times10^{-2}$ & $7.399\times10^{-4}$  \\
\mr
$4$ & $0.4637$  & $0.9779$  & $0.4782$             & $2.642$      & $11.049$   &$7.169\times10^{-2}$ & $5.046\times10^{-4}$\\
\br
\end{tabular}
\end{indented}
\label{tbl:IDBoBvalues}
\end{table}
\normalsize

The BoB model (as emphasized in \cite{arXiv:1810.00040}) does not depend directly on any other numerically fitted coefficients.
 We start building it now, by calculating the angular frequency of the fundamental mode using eq.\eref{eq:omgBoB}, with $\kappa$ given by eq.\eref{eq:constk} and $t_i$ from eq.\eref{eq:tisimp}, then we obtain the dominant mode by multiplying it with $m=2$. 
Once the frequency is known, we calculate the amplitude, given by eq.\eref{eq:ampBoB}. 
We proceed next to calculate the phase, given by eq.\eref{eq:phiBoB}.
Lastly, we implement the strain, given by \eref{eq:strainBoB}.
We remind the reader that we assumed the black holes entering the merger do not spin, or their effective spin is zero, and indeed this is a good approximation for most of the binaries detected in the LIGO and Virgo O1–O3 observing runs \cite{arXiv:1912.11716}.  

\subsection{Fasten It Tight}
\label{ssec:hybrid}

It is due time now to attach the GW template for the inspiral to the one obtained with the BoB mode and to construct a completely analytical GW template that encompasses the whole evolution of the BBH collision, from the time it enters the detection band untill the quiescence of the final BH.

We use an equal-mass binary of normalized mass $m_1+m_2=1M$ and start the inspiral at a separation $r_0=10M$, but those assumptions can be easily modified to fit any separation, and rescaled to a desired total mass of the BBH system, as well as mass ratio up to $q = 10$.  
For the initial eccentricity we will consider two values, $e_{i,1} = 10^{-3}$ corresponding to the quasi-circular case, and an eccentric close orbit of $e_{i,2}=0.15$. 
With this initial data we run the PN portion of the binary evolution. 
The  quasi-circular case evolves up to $T_{s,1} = 530 M$, its eccentricity drops to $e_{f,1}=1.7\times 10^{-4}$ and its orbital frequency
reaches  $\Omega_{f,1} = 0.13407$.
We take the initial frequency for the BoB model at $t_{LR,1}=T_{s,1} - 20M$ corresponding to the location of the LR, where the PN approximation breaks. The value of this frequency, taken directly from the PN data, is $\Omega_{i,1} = 7.4085\times 10^{-2}$. 
The large eccentricity case has a shorter span, only up to $T_{s,2} = 460 M$, its eccentricity drops to $e_{f,2}=1.988\times 10^{-2}$  and its orbital frequency raises to $\Omega_{f,2} = 0.19466$.
Even for this case, the initial frequency for the BoB model at $t_{LR,2}=T_{f,2} - 20M$  is $\Omega_{i,2} = 7.5837\times 10^{-2}$, which is close to the values found in Table \ref{tbl:LRvalues}. 
Note that the orbital frequency at the LR stays $\approx 0.075$ with a mean deviation of $\approx 10^{-4}$, marking the transition from the weak (far field) to the strong (near field) zone where the reduced wavelength of the gravitational wave $\bar{\lambda}_{GW} = \Omega^{-1}$ becomes comparable to the binary separation $r$.

Save for the slight difference in the initial frequency, we start the BoB model with the same values for the ringdown frequency and quality factor. 
We perform the stitching in frequency at the initial time calculated in eq.\eref{eq:tisimp} $t_i$ for the BoB model.

\begin{figure}[!ht]
\centering
\includegraphics[scale=0.75]{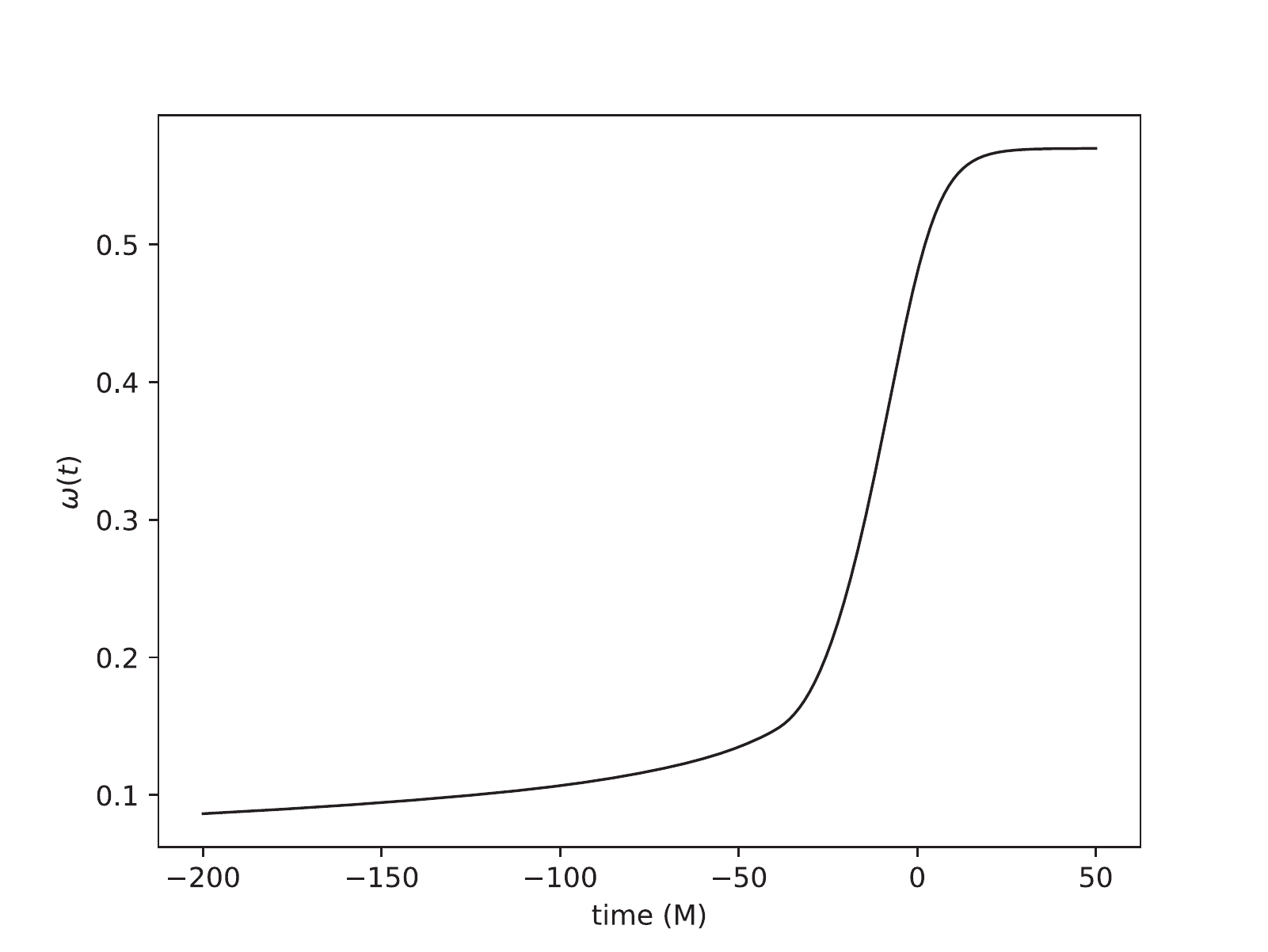}
\caption{Hybrid frequency, by matching the PN at $t_{LR,1}$ with BoB at $t_i$, for an equal-mass binary merger with initial eccentricity $e_i= 10^{-3}$ at $r=10M$.}
\label{fig:whyb}
\end{figure}

\begin{figure}[!ht]
\centering
\includegraphics[scale=0.75]{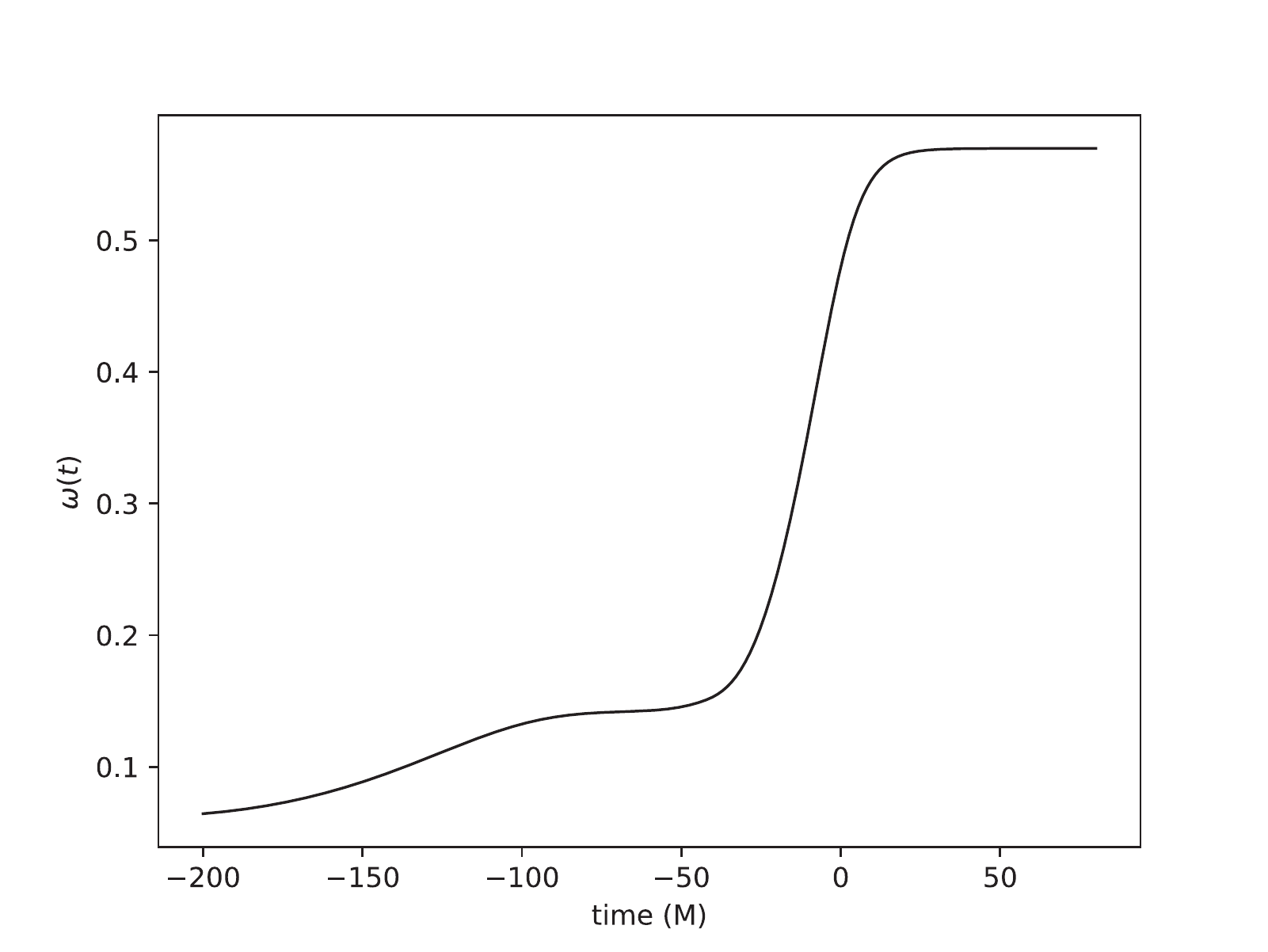}
\caption{Hybrid frequency, by matching the PN at $t_{LR,2}$ with BoB at $t_i$, for an equal-mass binary merger with initial eccentricity $e_i= 0.15$ at $r=10M$.}
\label{fig:whybe015}
\end{figure}

We create the hybrid frequency with the formula:
\begin{equation}
\omega_{hyb}(t)=(1 - \sigma(t)) \omega_{PN}(t - ti) + \sigma(t)\omega_{BoB}(t).
\end{equation}
where $\sigma(t)$ is the piecewise function:
\begin{equation}
\sigma(t)= \left\{
  \begin{array}{lr} 
      0 & t\leq t_i \\
      1 & t > t_i 
      \end{array}
\right.
\end{equation}
We obtain an excellent overlap at and around $t_i$, as shown in Fig.\ref{fig:whyb}, without introducing any other time shift. 
Even the high-eccentricity case, as shown in Fig.\ref{fig:whybe015}, exhibits this excellent behavior at the chosen matching time. 
We drastically reduced the ambiguity in time for the matching in frequency, which is a remarkable result. 

Before stitching together the strain, we normalize the BoB amplitude and rescale the amplitude of the inspiral with this normalized amplitude divided by the PN amplitude at the light ring $t_{LR}$.

\begin{figure}[!ht]
\centering
\includegraphics[scale=0.75]{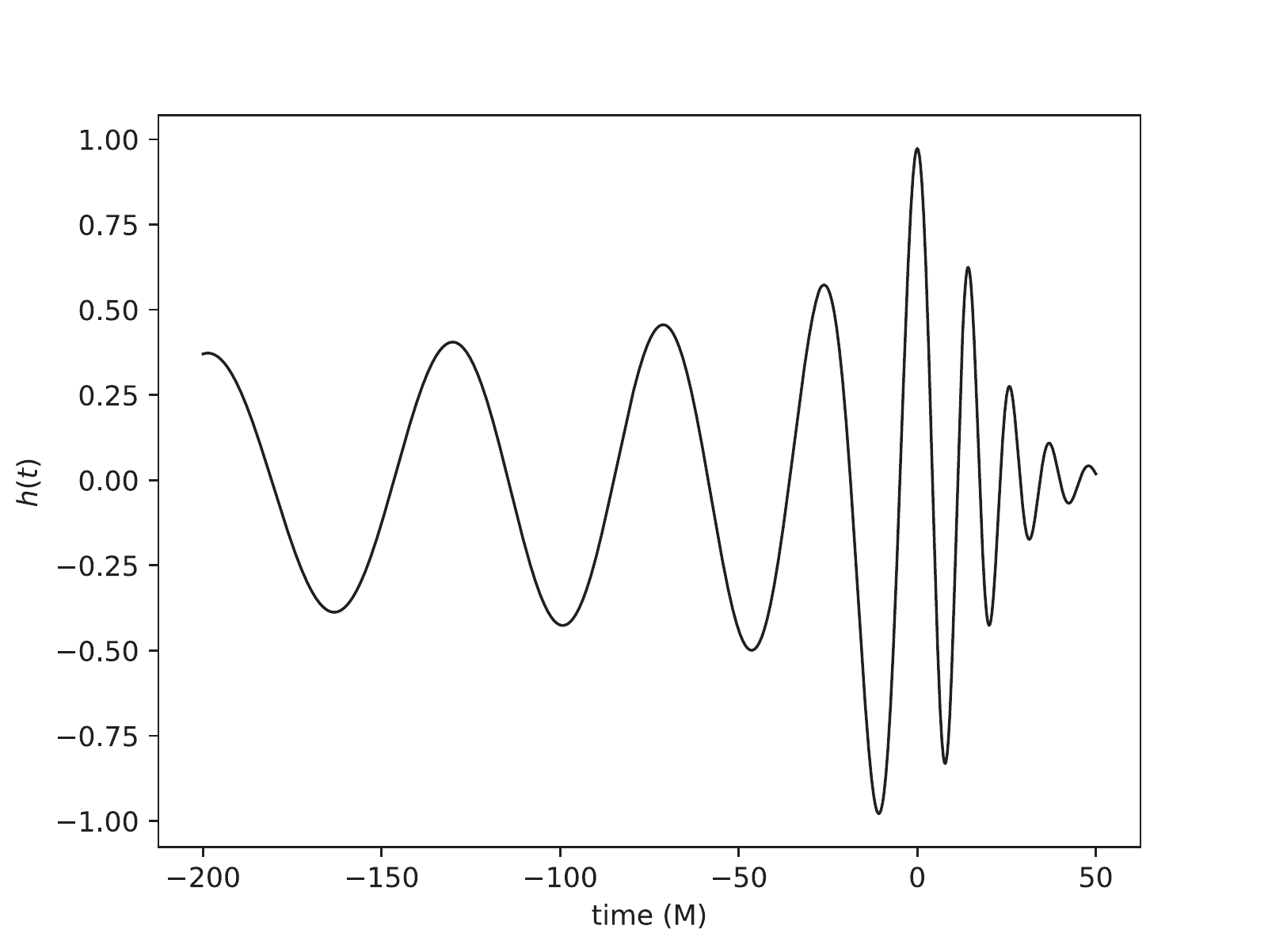}
\caption{Hybrid strain, by matching the PN at $t_{LR,1}$ with BoB at $t_i$, translated with $t_p$, for an equal-mass binary merger with initial eccentricity $e_i= 10^{-3}$ at $r=10M$.}
\label{fig:hMatched}
\end{figure}

\begin{figure}[!ht]
\centering
\includegraphics[scale=0.75]{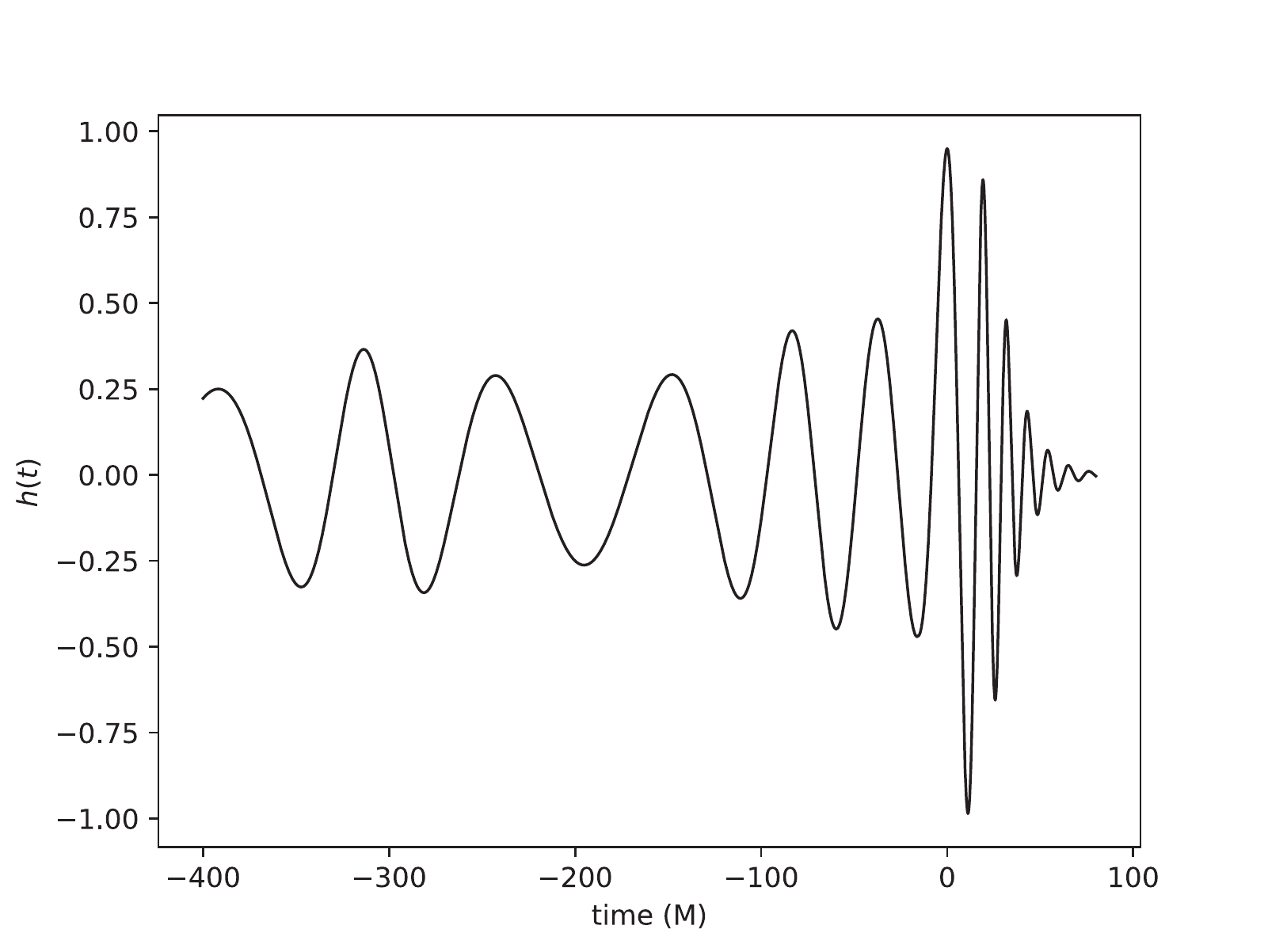}
\caption{Hybrid strain, by matching the PN at $t_{LR,2}$ with BoB at $t_i$, translated with $t_p/3$, for an equal-mass binary merger with initial eccentricity $e_i= 0.15$ at $r=10M$.}
\label{fig:hMatchede015}
\end{figure}

We form the hybrid strain with the same technique, this time translating also the BoB strain with the peak time of increase in frequency $t_f$ and correcting it for sign, which indicates a phase difference of $\pi/2$. 
We note that this does not introduce a shift ambiguity because the two models might not to use the same sign convention. The formula for the normalized hybrid strain is:
\begin{equation}
\bar h_{hyb}(t)=(1 - \sigma(t)) \bar h_{PN}(t - ti) - \sigma(t) \bar h_{BoB}(t+t_p).
\end{equation}
For the low-eccentricity case this provides again an excellent overlap at and around $t_i$, as seen in Fig.\ref{fig:hMatched}.   
When matching the high-eccentricity inspiral with the BoB merger strain we encounter a slight mismatch, both in amplitude and time of the matching. This is most likely due to the fact that the orbit is not fully circularized by the time the binary enters the strong-field zone, and the BoB model does not take into account the eccentricity. 
We mitigate residual effects or the eccentricity by matching the two strains at the closest peak in their amplitude near $t_i$ and rescaling slightly the inspiral strain to match the amplitude of the BoB at that time.
The result is shown in Fig.\ref{fig:hMatchede015}. This indeed introduces an expected ambiguity in the matching interval for high-eccentricities, which can be mitigated in further studies.          

\section{Conclusions}
\label{sec:conclude}

In this work we implemented a purely analytical framework for obtaining complete GW templates from eccentric binary sources and produced a set of new, freely available \texttt{Wolfram Mathematica} notebooks.
First, we gathered and assembled into a coherent way all the pieces of the puzzle required to ensure the accurate implementation.
We started with an exposition of the essential theoretical framework, by introducing the specific terminology and by giving clear explanations to prevent confusing readers with different scientific background.
We carefully chose, from the rich scientific literature, a fully analytical procedure to build the inspiral GW that contains high-order PN corrections for enhanced accuracy. 
For the eccentricity and slowness we employ instantaneous and hereditary coefficients with energy corrections up to $6$PN. 
We implemented a purely analytical expression for the solution to Kepler's equation that gives the eccentric anomaly, accurate up to $3$PN order. 
We calculated the orbital separation and the phase, keeping the errors below machine precision, and used high order interpolating polynomials when necessary to speed up the calculation. We assembled the strain and displayed its dependence on eccentricity and mass ratio, showing that we expect a higher number of GW detections from eccentric comparable mass ratios binaries, which emphasizes the relevance of modeling such systems.
For the merger, we implemented the purely analytical BoB model for the merger, which relies on the quasi-circular assumption.
At last, we glued together the eccentric inspiral and the quasi-circular merger waveforms by completing them in frequency, amplitude and phase with a piecewise function, building the hybrid GW strain for the whole evolution of the binary. 
Remarkably, for low eccentricity, the match between the BoB and PN approximations are on-point, with coincidence in the overlap, which indicates no ambiguity in the time interval, a great improvement from the usual matching techniques.
For high-eccentricity we must make a slight adjustment, both in time and amplitude, to compensate for the implicit quasi-circular assumption built into the BoB approach, but the matching in frequency and phase are again unambiguous.

The new open-source tool for the calculation of analytic GW templates for eccentric binaries, together with the thorough and streamlined documentation of our steps given in this paper, offers researchers in the field of gravitational waves a straightforward path to understand, reproduce, use and extend our implementation.  
Future developments will incorporate higher PN orders, will include spin, precession and kicks, and might augment the BoB model to account for eccentricity close to the merger. Our notebooks are available at \texttt{github.com/mbabiuc/MathScripts}. 

\ack
The authors wish to acknowledge assistance from the Physics Department at Marshall University, and financial support from the National Science Foundation through the EPSCoR Grant OIA-1458952 to the state of West Virginia “Waves of the Future”.

\section*{References}

\bibliography{references.bib}

\end{document}